\documentclass[prb,twocolumn,showpacs]{revtex4}\usepackage{graphicx}
\begin{document}
\title{Exciton condensation in quantum wells. Excitonic hydrodynamics. Autosolitons}
\author{V.I.~Sugakov}
\altaffiliation{Institute for Nuclear Research,National Academy
of Sciences  Kyiv,Ukraine)}\email{sugakov@kinr.kiev.ua}
Comments: 11 pages, 8 figures
Subjects: Mesoscale and Nanoscale Physics (cond-mat.mes-hall)
\begin{abstract}
The hydrodynamics equations for indirect excitons in the double
quantum wells are obtained taking into account 1) a possibility of
an exciton condensed phase formation, 2) the presence of a
pumping, 3) the finite value of the exciton lifetime, 4) the
exciton scattering by defects.
  New types of  solutions in the form of bright
and dark autsolitons are obtained for the exciton density.  The
role of localized and free exciton states is analyzed in a
formation of the emission spectra.
\end{abstract}
\maketitle

\section{Introduction} While seeking  the exciton Bose-Einstain condensation in  double quantum wells, the several
 non-trivial effects were observed. A very large exciton lifetime is a special feature of excitons in double quantum well
 in the presence of an electric field directed in parallel to the normal of quantum well plain \cite{1}. The effect occurs due
 to the separation of electrons and holes into the different wells, which causes a very weak overlapping their wave function and
  the damping of the mutual recombination. The large lifetime allows one to create a high  concentration of the excitons  at small pumping and
  to study a manifestation of the effects of exciton-exciton interaction. Excitons with electrons and holes localized in different wells are
  called 'indirect excitons'. The nonzero dipole moment of the indirect excitons should cause their mutual repulsion
  that complicates the creation of the exciton condensed phase. These properties and the facts, that excitons have the integer
  spin and the small mass, stimulated the search  of the exciton Bose-Einstein condensation  in double quantum wells.
  These investigations gave a number of new results. So, the narrow band with unusual properties was observed and studied by Timofeev's
   group\cite{2,3,4} in the emission spectra of the indirect excitons in $AlGaAs$ system. It was shown
   that this band appears at the some threshold pumping, also the peculiarities
   of the  temperature and pumping dependencies  were revealed.  The authors \cite{3,4} builded a
      phase diagram "threshold pumping - temperature".

The nontrivial results were found in a spatial distribution of
exciton emission from quantum well. In the papers \cite{5}, an
appearance of a ring outside the laser spot was observed in the
emission from double quantum well.  The ring radius  exceeded
significantly the exciton diffusion length. The explanation of the
appearance of the ring was given in the papers
 \cite{7,8} under the assumption, that holes are captured by the well more effectively than electrons,
 and, in addition,  there are donors in the crystal, which create some concentration of free electrons.
  As a result a region rich by holes arises in a vicinity of the laser spot.
   Outside of this region the quantum well is enriched by free electrons. On the boundary of the region
   the processes of a recombination take place, which  causes the a creation of excitons on the ring
   and the appearance of a spatial distribution of the emission in the form of the ring.

Intriguing facts appear under an investigation of the spatial
distribution of the exciton density. Different spatial
nonhomogeneous structures  were observed in the emission of the
indirect excitons at the pumping greater a critical value. Thus,
in the paper  \cite{5} a division  of the emission ring was
observed on separate fragments periodically localized along the
ring.  In the paper \cite{9}, in which the excitation of the
quantum well carried out through a window in metallic electrode,
the authors found  in the luminescence spectrum  a periodical
structure of the islands situated along the ring under perimeter
of the window. The number of the islands growths with increasing
of the radius of the window. The similar results were obtained not
only in double quantum wells, but also in a wide quantum well in
the presence of a strong electric field, which  divides electrons
and holes between different sides of the quantum well. As the
result the excitons with  charges, strong separated in the space,
and with large lifetime are created \cite{10}. On such dipole
excitons in the wide quantum well, the effects  similar to those
 on  excitons in double quantum wells were observed,namely the
appearance of periodical structures of the islands under the
window in a electrode\cite{11} and on the ring outside the laser
spot \cite{All}. Recently Timofeev and coauthors \cite{12,Gorb}
presented examples of the structures in the emission spectra  at
the different forms of windows in the electrode: the rectangle,
two circles and others. In the paper \cite{13} the authors,
choosing the form of the electrode, created additional periodical
potential for excitons. It  was found, that besides the periodical
structure imposed by external conditions, the partition of the
emission into fragments was observed in the direction, in which
the potential is almost uniform.

The phenomenon of the symmetry loss and the  creation of
structures in the emission spectra of indirect excitons stimulated
a series of theoretical investigations
\cite{14,15,16,17,18,18a,18b,18d}. The authors of the work
\cite{14} considered the instability, which arises under kinetics
of  level occupations   by the particles with the Bose-Einstein
statistics. Namely, the growth of the occupation of the level with
zero moment should stimulate the transitions of excitons to this
level. But the density of excitons was found greater, and the
temperature was found lower than these values observed on the
experiments. Some authors explain the appearance of the
periodicity by Bose condensation of excitons \cite{15,16}. There
is a suggestion   to describe the system by a nonlinear
Schrodinger equation  \cite{17,18}. Also  a possibility of the
Mott transition in  considered systems was studied \cite{18a}. The
periodicity appearance in the exciton system was investigated in
\cite{18b} using Bogolyubov's equations under some approximations
of inter-exciton interaction. Let note the paper \cite{18d},in
which the observed in \cite{5} appearance of the islands was
explained from classical model diffusion equation with Coulomb
interaction between electrons,holes, and excitons. In the listed
above works the main efforts were applied for the ascertainment of
the principal possibility of the appearance of  the periodicity of
the exciton density distribution. A specific application of the
results for an explanation of the numerous experiments of
different kind (at different pumping and temperature,
nonhomogeneous external fields) was not employed.

  An appearance of an instability of the uniform distribution of the exciton density and a
  formation  of the
periodical distribution were shown in the work
 \cite{18e} from a position of self-organization processes in non-equilibrium systems for  excitons with attractive
 interaction.  The phenomenon has the threshold behavior with respect to a pumping.
    After the successful observation  of periodical structures
 in the system of indirect excitons in double quantum well on the base of AlGaAs crystal
  by Timofeev's and Butov's groups, the several
 works were fulfilled \cite{19,20,21,22,23,23a,23b} in this model
 devoted to explanation of the experiments.
  The theoretical approaches of the  works \cite{19,20,21,22,23,23a,23b}
are based on the following assumptions.

1.There is an exciton condensed phase caused by the attractive
interaction  between excitons. As was mentioned, there are the
dipole-dipole repulsion interaction between excitons.  But the
simple calculations show, that the exchange and van der Waals
interactions exceed the dipole-dipole repulsion at certain
distances between excitons, if the distance between quantum wells
is   not very far and the exciton dipole moment is not too large.
 An
existence  of attractive interaction between excitons is confirmed
by the calculations of biexcitons  \cite{24,25,26,26a}, and  under
investigation of  many-exciton system \cite{27}.

2.The finite value of the exciton lifetime plays an important role
in the formation of a spatial distribution of exciton condensed
phases.  As usually, the exciton lifetime exceeds significantly
the duration  of the establishment of a local equilibrium. By this
reason, the lifetime of excitons is suggested to be equal infinity
under the solutions of many exciton problems. But, the taking into
account the finiteness of the exciton lifetime is necessary in the
study of the spatial distribution phases in two-phase systems,
because the exciton lifetime is less than the time of the
establishment of the equilibrium between phases. The last time is
determined by slow diffusion processes and  is great. Just the
finite exciton lifetime restricts the maximal size of the exciton
condensed phase and it causes the existence of a correlation in
positions of separate regions of the condensed phase. As the
result the spatial structures of the condensed phase appear in the
shape of separate islands (in two dimensional case), parameters of
which and the mutual position depend on the exciton lifetime.
Thus, the created spatial structures are non-equilibrium and they
are a consequence of a self-organization in non-equilibrium
systems.

The theory, developed in the works \cite{19,20,21,22,23,23a,23b},
has explained many  features of the indirect exciton manifestation
in the double quantum wells on the base $AlGaAs$ crystal. So, the
observed in \cite{4} behavior of narrow band as a function of the
pumping and the temperature is presented in \cite{20}. It is
follow from these calculations in accordance
  with the experiment \cite{4}, that the intensity of the emission of the narrow
band
 decreases  linearly  with temperature  at fixed pumping and rises ultralinearly with the growth of the pumping at fixed
 temperature . The
theory \cite{20} has explained the experimental "threshold pumping
- temperature"  phase diagram obtained in \cite{3,5}. The works
\cite{19,21,22,23} were dedicated to the interpretation of the
appearance of the periodically situated islands in the emission of
spectra from the both the ring outside the laser spot, founded in
\cite{5}, and  from the region of the double quantum well under
perimeter of the window in electrode, founded in \cite{9}. The
theory described the sizes of the islands, the distances between
them, their appearance and vanishing depending on  pumping and
temperature. The calculated behavior of the condensed phase
islands around  two laser spots observed in \cite{8} is presented
in \cite{23b}. At approaching the centers of lasers spots, the
rings of the emission around two spots transform from two rings
around two centers to the deformed single ring with two laser
spots inside the ring in according with experiment \cite{8} . The
theory \cite{20} explained an appearance of a spike observed in
\cite{But}
  in emission of the indirect excitons after the shutdown of a
  pumping  by the increasing of
  the exciton lifetime in a consequence of the removal of an Auger
  processes. In the paper\cite{22} the fragmentation of the inner
  ring in the laser spot was predicted. These
  fragmentation was observed recently in the experiment \cite{Rem}.

   Several theoretical investigations were made for new systems, which were not
  studied  by experiments so far. It was shown, that a periodical
  structure of the condensed phase islands arise at the light
   irradiation in the quantum
  well under electrode with a slot \cite{Sug}. The chain of the
  islands moves in the presence of a linear potential along the
  slot. The process reminds the Gunn effect in semiconductors and can be called "excitonic Gunn effect".
   The pumping dependence study of exciton density
  distribution
  in the well under the electrode in the shape of a disk
  presents all stages of the phase transitions: from the islands of
  the condensed phase in a gas phase till the islands of the gas
  phase in an environment of the condensed phase \cite{23c}. The
  investigation of excitonic pulse moving in external fields
  \cite{23d} shown that maximum exciton density remains constant during the exciton lifetime and
  there is a possibility of  a control of pulse moving by an another laser
  if the formation of the pulse occurs by excitons in the
  condensed phase.

 While  developing the theory, two
approaches of the theory of phase transitions were used: the model
of nucleation (Lifshits-Slyozov) and the model of spinodal
decomposition ( Cahn-Hillart). These models were generalized on
the particles with the finite  lifetime, that is important
 for interpretation of the experimental results.  The involvement of Bose-Einstein
 condensation for excitons was
  not required for the explanation of the experiments,
  the considered condensation is the condensation in real space.
   Among many experiments, explained by the theory,
  there are two ones, presented below, which did not considered jet in the framework of the presented theory.
\newline \indent
 A. It was shown in the work  \cite{30}, that
maximum of the frequency dependence of the emission from the
region between the islands is lower, than maximum of the emission
frequency from the islands, so from the region, where the exciton
density is large. The difference of the maxima  is small, it is
less than the width of the emission band.
  But on the base of this date the authors came to the conclusion, that there is the repulsion interaction between excitons only.
  This result contradicts the main assumptions of the model of the works
 \cite{19,20,21,22,23,23a} about the presence of attractive interaction between
 excitons, that exist at some distance between excitons and
  causes  the creation of a condensed phase. In the Appendix we
 present a possible explanation of such effect
 in the case of the  attractive interaction between  excitons \cite{SugUJP}. The explanation is based
 on the presence in the well of localized exciton states, levels of which are situated lower than the exciton
 levels and become saturated with increasing pumping. The emission band is determined by the free and localized exciton states.
 The exciton states form the upper part of the band. With increasing pumping the number of excitons and
 the blue part of the band emission growth.
  In the case of an attractive interaction, a lowering of the exciton levels caused by the exciton condensation  is
 small in comparison with the whole width of the band.  As the
 result  the maximum of the emission shifts to higher side with increasing the pumping, if
 the emissions from the  exciton condensed phase and from the localized states
 are not separated.
\newline \indent
 B. In the works \cite{12,31,32} a coherence was observed in the
emission spectra from island  \cite{32}, or even from different
islands \cite{12,31}. The coherence was revealed  in the
interference of the emission from the different spatial points.
\newline \indent
The effect $B$ is not considered in the presented paper. For its
study, the microscopic model of the condensed phase is needed. But
a qualitative explanation may be given.  The interference of the
wave functions  does not observed on the experiment directly, the
interference of  electromagnetic waves is shown on the experiment.
Because the electromagnetic field and scattered field are
coherent, the interference of the emission from two islands may
arise as a result of an imposition of electromagnetic field
emitted by some island and the scattered  field by other island.
It was shown in the papers \cite{33,34,35}, that the strong
correlation between exciton densities at different points takes
place in the case when the exciton condensed phase exists. Also
there is a sharp maximum of the Fourier transformation of the two
point correlation function.
 It is reason of a mutual connection of the wave emitted from
  some point with the wave scattered by other region.
 But the quantitative calculations require the date of
microscopic model of the exciton condensed phase, particularly,
the numerical value of the polarizability  is needed.
\newline \indent
 In the presented paper the hydrodynamics equation for excitons is
obtained for the case, if excitons are in condensed phase. The
equations allow to describe the moving of the complicated system
composed of two phases: gas and condensed ones. We have studied
the spatial distribution of both the exciton density and exciton
flux in the case of condensation at steady-state pumping.  From
analysis of different solution of the equation for exciton density
it was shown the existence of exciton autosolitons at some
parameters of the system. Also the possible explanation of the
effect $A$ is given
 taking into account  the presence of localized
states, which  become totaly occupied  with increasing pumping.

It is necessary to emphasize that besides the  system, we
investigated, there is an another one, in which the exciton
condensation is studied. It is the exciton condensation in bilayer
quantum Hall system. In this system the two layers are filled by
electrons in the presence of strong magnetic field with total (for
two wells) Landau level filling factor $\nu_T =1$ and with total
density $n_T=n_1+n_2$, where $n_i$ is the density in the separate
well ($i=1,2$). An electron in the lowest Landau level of one
layer bounded to hole in the lowest Landau level of the other
layer is considered as exciton. The review of collective effects
in such system is presented in\cite{Eis} . The bilayer quantum
Hall system differs from the system that we consider.  The
electrons are created in the bilayer quantum Hall system  by
donors and the system is equilibrium. We studied the system in
which the excitons are created by the light,the system is
non-equilibrium. The excitons have the finite value of the
lifetime and this fact influences significantly on the behavior of
the collective states, particularly, on the formation of the
spatial structures.

\section{Hydrodynamics of exciton condensed phase}

The hydrodynamic equations of excitons were obtained and analyzed
in the work \cite{36}. In comparison with this paper, we  obtain
\cite{SugUJP} the hydrodynamic  equations of excitons
 generalizing the Navier-Stokes equations taking into account the finite exciton lifetime, the pumping
  of exciton and the  existence of an exciton condensed phase caused by interaction between excitons.
  The system is described by the exciton density   $n \equiv n(\vec {r},t)$ and by the velocity of the exciton liquid $\vec
{u} \equiv \vec {u}(\vec {r},t)$. The equation of the continuity of the exciton density is rewritten in the form
\begin{equation}
\label{eq8}
\frac{\partial n}{\partial t} + div(n\vec {u}) = G - \frac{n}{\tau _{ex} },
\end{equation}
where $G$ is the pumping (the number of excitons created for unit
 time in unit area of the quantum well), $\tau_{ex}$ is the
 exciton lifetime.
In the comparison with the typical equation for a liquid, the
presented equation for excitons contains the terms, that describe
the pumping and the finite lifetime of the excitons.

The equation for the movement of the unit volume of an exciton liquid is rewritten in the form
\begin{equation}
\label{eq9}
\frac{\partial mnu_i }{\partial t} = - \frac{\partial \Pi _{ik} }{\partial
x_k } - \frac{mnu_i }{\tau _{sc} },
\end{equation}

\noindent
where  $m$ is the exciton mass, $\Pi _{ik} $ is the tensor of the density of the exciton flux.
\begin{equation}
\label{eq10}
\Pi _{ik} = P_{ik} + mnu_i u_k - {\sigma }'_{ik} ,
\end{equation}
\noindent
where
 $P_{ik}$  is the pressure tensor,
${\sigma }'_{ik}$ is the viscosity tensor of a tension.

In comparison with the typical Navier-Stokes equation a braking of
exciton liquid by phonons and by defects is introduced in Eq.
(\ref{eq9}).  In the equation (\ref{eq9}) we neglected by the
momentum change caused by the creation and the annihilation of
excitons. Indeed, the momentum change in the unit time and in the
unit volume owing to an disappearance of the excitons has  the
order $mnu / \tau _{ex} $. Since  $\tau _{ex}
> > \tau _{sc} $ this value is much less the last term in the
formula  (\ref{eq9}). The momentum change due to  the addition of
new excitons by pumping is small too, because mean exciton
 moment, created by external light, is close to zero.

 Introducing coefficients of the viscosity  and using Eq.
(\ref{eq8}), Eq.(\ref{eq9}) may be rewritten in the form
\begin{eqnarray}
\label{eq11} \rho \left( {\frac{\partial u_i }{\partial t} +
\left( {u_k \frac{\partial }{\partial x_k }} \right)u_i } \right)
= - \frac{\partial P_{ik} }{\partial x_k } + \eta \Delta u_i
\nonumber\\+ (\varsigma + \eta / 3)\left( {\frac{\partial
}{\partial x_i }} \right)div\vec {u} - \frac{\rho u_i }{\tau _{sc}
},
\end{eqnarray}
$\rho=mn$  is the mass of excitons in unit volume.

Let us consider the tensor of pressure. To find the connection
between the tensor and others parameters it is needed to use the
equation of the state. We suggest that the state of the local
equilibrium is realized and the state of the system may be
described by a free energy, which depends on spatial coordinate.
Let us present the functional of the free energy in the form
\begin{equation}
\label{eq12}
F = \int {d\vec {r}\left( {\frac{K}{2}(\nabla n)^2 + f(n)} \right)} .
\end{equation}

The first term in the integrand  (\ref{eq12}) describes the energy
of non-homogeneity.

The pressure tensor will be obtained from equation of the state of
the system. At the given presentation of the free energy, the
pressure tensor is determined by the formula \cite{37}
\begin{equation}
\label{eq13} P_{\alpha \beta } = \left( {p - \frac{K}{2}(\vec
{\nabla }n)^2 - Kn\Delta n} \right)\delta _{\alpha \beta } +
K\frac{\partial n}{\partial x_\alpha }\frac{\partial n}{\partial
x_\beta },
\end{equation}
\noindent where $p = n{f}'(n) - f(n)$ is the equation of the
 state, $p$ is the isotropic pressure.

Taking into account  (\ref{eq13}), we rewrite the equation
(\ref{eq11}) finally  in the form \cite{SugUJP}
\begin{eqnarray}
\label{eq14}
\frac{\partial u_i }{\partial t} + u_k \frac{\partial u_i }{\partial x_k } +
\frac{1}{m}\frac{\partial }{\partial x_i }\left( { - K\Delta n +
\frac{\partial f}{\partial n}} \right) + \nu \Delta u_i
\nonumber\\+ (\varsigma / m +
\nu / 3)\left( {\frac{\partial }{\partial x_i }} \right)div\vec {u} +
\frac{u_i }{\tau _{sc} } = 0.
\end{eqnarray}

\noindent Eqs.(\ref{eq8},\ref{eq14}) are the equations of
hydrodynamics for an exciton system. These equations differ from
the hydrodynamic equations, investigated in  \cite{36}, by the
presence of the terms of the right side in Eq.(\ref{eq8}), which
take into account the lifetime and pumping,  and  by the presence
of the third term in  Eq.(\ref{eq14}), which describes a condensed
phase. It follows from the estimations, made in the work
\cite{36},
 that the terms with the viscosity coefficients are small and we shall neglect them.

 In the case of a steady state irradiation of the system, the
 Eqs.(\ref{eq8})
 and (\ref{eq14}) have the solution $n=G\tau$, $u=0$. In order to
 investigate the stability of this solution we consider the
 behavior of a small fluctuation of the exciton density and the
 velocity from these values: $n\rightarrow n+\delta n\exp(i\vec{k}\cdot\vec{r}+\lambda
 t)$, $u=\delta u\exp(i\vec{k}\cdot\vec{r}+\lambda(\vec{k})
 t)$.  After substitution  these expressions in Eqs.(\ref{eq8},
 \ref{eq14}), we obtain in the linear approximation with respect to the
 fluctuations the following expression
\begin{eqnarray} \label{eq14aa} \lambda_{\pm}(\vec{k})=\frac{1}{2}(-(1/\tau_{sc}+1/\tau_{ex})
\nonumber\\\pm\sqrt{(1/\tau_{sc}-1/\tau_{ex})^2-
\frac{4k^2n}{m}(k^2K+\frac{\partial^2 f}{\partial n^2})} ),
\end{eqnarray}

It is follow from (\ref{eq14aa}), that both parameters
$\lambda_{\pm}(\vec{k})$ have  a negative real part at small and
large values of vector $\vec{k}$ and, therefore, the uniform
solution of hydrodynamics equation is stable. But the value
$\lambda_{+}(\vec{k})$ may be positive in some interval of vector
$\vec{k}$ , when $\frac{\partial^2 f}{\partial n^2}$ becomes
negative. In these case the uniform distribution of the exciton
density is unstable with respect to a formation of nonhomogeneous
structures. The instability arises at some threshold value of
exciton density and at some  critical value of the wave vector.
Analysis of the equation (\ref{eq14aa}) gives the following
expression for the critical values of the wave vector $k_c$ and
the exciton density $n_c$
\begin{equation} \label{eq14ab}
k_c^4=\frac{m}{Kn_c\tau_{sc}\tau_{ex}},
\end{equation}
\begin{equation} \label{eq14ac}
\frac{k_c^2n_c}{m}\left(k_c^2K+\frac{\partial f(n_c)}{\partial
n_c^2}\right)+\frac{1}{\tau_{sc}\tau_{ex}}=0.
\end{equation}

For stable particles ($\tau_{ex}\rightarrow\infty$) the equations
(\ref{eq14ab},\ref{eq14ac}) give the condition $\frac{\partial
f}{\partial n^2}=0$, that is condition for spinodal decomposition
for a system in the equilibrium case.

Depending on parameters the Eqs.(\ref{eq8},\ref{eq14}) describe
the ballistic  and diffusion movement of the exciton system.  The
relaxation time $\tau _{sc}$ plays the important role in a
formation of the exciton moving.  Due to arising of the
nonhomogeneous structures, the exciton currents  in the system
exist  ($\vec{j}=n\vec{u}\neq 0$) even under
 the uniform steady-state pumping. Excitons are moving from  regions with the small exciton density
 to the regions with the high density. In the presented paper we shall consider
the spatial distribution of exciton density and exciton current in
the double quantum well under steady-state pumping.   In this case
the exciton carrent is small and we suggest the existence of the
next conditions
\begin{equation}
\label{eq14a} \frac{\partial u_i }{\partial t}<<u_i/\tau _{sc}.
\end{equation}
 \begin{equation}
\label{eq14b} u_k \frac{\partial u_i }{\partial x_k }<< u_i /\tau
_{sc}.
\end{equation}
Particularly, the Eq.(\ref{eq14a}) holds under the study of
the steady-state exciton distribution. The fulfilment of Eq.(\ref{eq14b}) will be shown later after some numerical calculations.

 Using  the conditions (\ref{eq14a}) and (\ref{eq14b})
   we obtain from Eq.(\ref{eq14}) the value of the
velocity  $\vec {u}$
 \begin{equation}
\label{eq15} \vec {u} = - \frac{\tau _{sc}}{m}\vec {\nabla }\left(
{ - K\Delta n + \frac{\partial f}{\partial n}} \right),
\end{equation}
  As the result  the equation for the exciton density current  may be presented in the form
\begin{equation}
\label{eq16} \vec {j} = n\vec {u}=- M\nabla \mu,
\end{equation}
\noindent  where $\mu = \delta F/\delta n$ is the chemical
potential of the system,
 $M = nD/\kappa T$ is the mobility, $D =
\kappa T\tau _{sc} /m$ is the diffusion coefficient of excitons.

Therefore, the equation for the exciton density (\ref{eq8}) equals
\begin{eqnarray}
\label{eq17} \frac{\partial n}{\partial t} = \frac{D}{\kappa T}( -
Kn\Delta ^2n - K\vec {\nabla }n \cdot \vec {\nabla }\Delta n)\nonumber\\
 +\frac{D}{\kappa T}\vec {\nabla }\cdot \left(
n\frac{\partial ^2f}{\partial n^2}\vec {\nabla }n \right)  + G -
\frac{n}{\tau _{ex} }.
\end{eqnarray}

Just in the form of (\ref{eq17})  we investigated a spatial
distribution of the exciton density at exciton condensation at
different dependencies  $f$ on $n$ [22, 23, 33, 38, 39]. So, our
previous consideration of the problem corresponds to the diffusion
movement of hydrodynamics equations (\ref{eq8},\ref{eq14}).
 At some conditions, applied to the functional $F$ the uniform solution is unstable,
 and the spatial structure arises in the system. For the  system
 under study
 the condensed phase appears, if  the function $f(n)$ describes a phase transition.
 In the papers mentioned above the examples of such dependencies were given.
Here, we  analyze an other dependence $f(n)$, which often is also
used in the theory of phase transitions. We  shall approximate the density of
the free energy in the form
\begin{equation}
\label{eq18}
f = \kappa Tn(\ln (n / n_a ) - 1) + a\frac{n^2}{2} + b\frac{n^4}{4} +
c\frac{n^6}{6},
\end{equation}
\noindent where $a$, $b$, $c$ are the constant values.  Three last
terms in the formula (\ref{eq18}) are the main terms, they arise
due to an exciton-exciton interaction and describe the phase
transition.
 The first term was introduced in order to describe the system
 in a space, where the exciton concentration
is small(if such region exists in the system). With increasing the
exciton density the term $a\frac{n^2}{2}$ must manifests itself
firstly. It gives the contribution the $an$ value   to chemical
potential. The origin of this term is connected in our system with
the dipole-dipole interaction that should become apparent at the
beginning due to its long-range nature. For estimations of $a$ we
may use for the dipole-dipole exciton interaction in double
quantum well the plate capacitor formula $an=4\pi e^2
dn/\epsilon$, where $d$ is the distance between wells, $\epsilon$
is the dielectric constant. This formula is used usually for a
determination of the exciton density from the the experimental
meaning of the blue shift of the frequency of the exciton emission
with the rise of the density. It is follow from the formula that
$a=4\pi e^2 d/\epsilon$.  This expression is approximate
 because it does not take into account the
exciton-exciton correlations \cite{25,37b}. When the exciton
density growths the last two terms in (\ref{eq16}) begin to play a
role. An existence of the condensed phase requires that the value
$b$ was negative($b<0$). For stability of the system at large $n$
the parameter $c$ should be positive ($c>0$). It is suggested in
the model, that the condensed phase arises due to the exchange and
van der Waals interactions.  For the system with the large
distance between wells the dipole-dipole interaction exceeds
attractive interaction and the condensed phase does not arise. The
disappearance of chemical potential minimum as a function of $n$
with increasing of the parameter $a$ (with increasing the distance
between the quantum wells) is presented in Fig1.
\begin{figure}\centerline{\includegraphics[width=8.6cm]{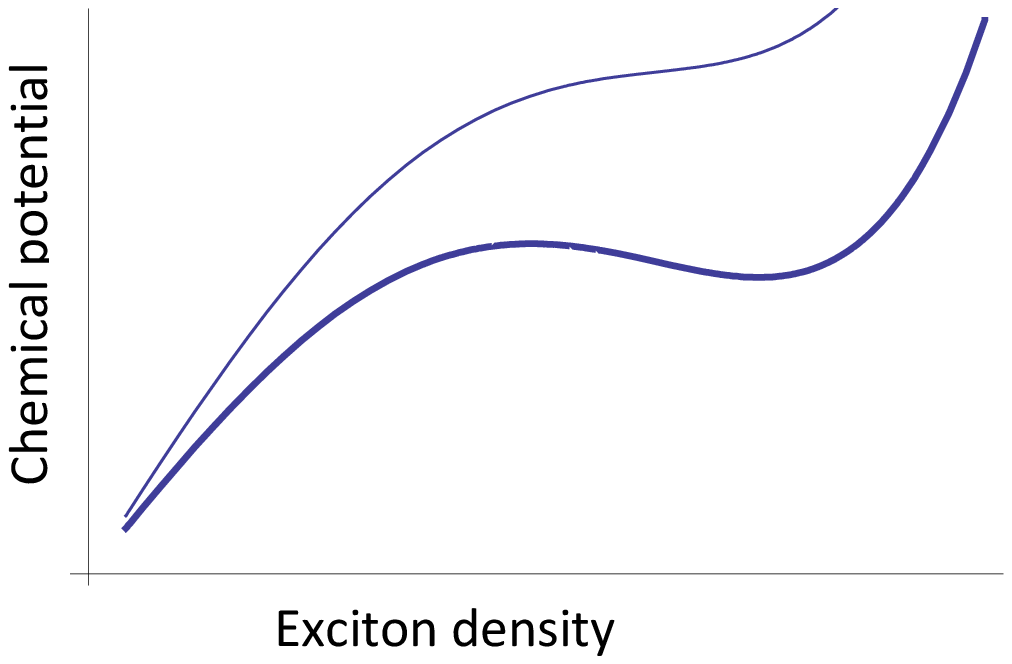}}\caption{
 Qualitative dependence
of the chemical potential on $n$.  The parameter $a$ for the thin
line is greater than for the thick line. The parameters $b$ and
$c$ are the same for both curves.}\end{figure}

 Let us introduce the dimensionless parameters:
$\tilde {n} = n / n_o ,$ where $n_o = \left( {a / c} \right)^{1 /
4}$, $\tilde {b} = b / (ac)^{1 / 2}$, $\tilde {\vec {r}} = \vec{r}
/ \xi $, where $\xi = \left( {K / a} \right)^{1 / 2}$ is the
coherence length , $\tilde {t} = t / t_0 $, where $t_0 =
\frac{\kappa TK}{Dn_o a^2}$, $D_1 = \frac{\kappa T}{an_o }$,
$\tilde {G} = Gt_0/n_0 $, $\tilde {\tau }_{ex} = \tau / t_0 $. As
the result the equation (\ref{eq17}) is reduced to the form
(hereinafter the symbol $\sim $ will be omitted in the equation)
\begin{eqnarray}
\label{eq19} \frac{\partial n}{\partial t} = D_1 \Delta n -
n\Delta ^2n + n\Delta n(1 + 3bn^2 + 5n^4)\nonumber \\- \vec
{\nabla }n\cdot\vec {\nabla }\Delta n+(\vec {\nabla }n)^2(1 +
9bn^2 + 25n^4) + G - \frac{n}{\tau _{ex}}.
\end{eqnarray}

  The  solutions of the equation
 (\ref{eq19})
  are presented in Fig.2 in the one-dimensional case ($n(\vec {r},t) \equiv n(z,t))$) for three values of
  the steady-state uniform pumping.
   \begin{figure}\centerline{\includegraphics[width=8.6cm]{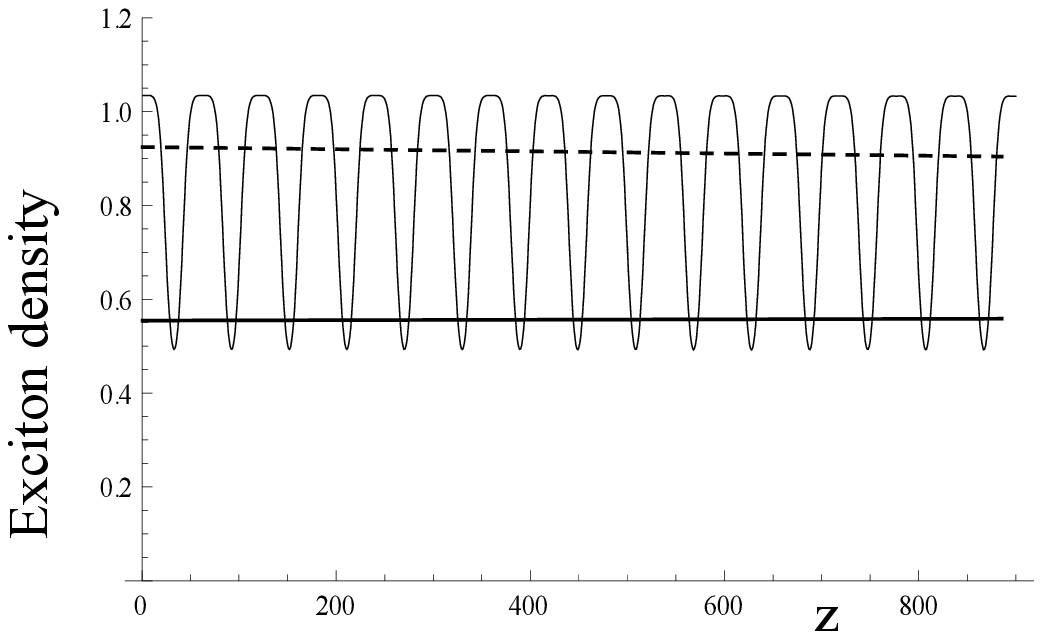}}\caption{
 The spatial dependence of the exciton density at the different value of the pumping:
 for the continues line   $G=0.0055$, for the periodical line $G=0.008$, for the dashed line
 $G=0.0092$. $D_1=0.03$, $b=-1.9$   }\end{figure}
The solutions are obtained  at the initial conditions  $n(z,0) =
0$ and the boundary conditions $n'(0,t) = n'(L,t) = n''(0,t) =
n''(L,t) = 0$, where $L$ is the size of the system.   The
periodical solution exists in the some interval of the pumping
$G_{c1} < G < G_{c2} $. At the great size  of the system the
structure of the solution (the period and the amplitude of the
lattice) does not depend on the boundary conditions.  At the given
parameters the periodical solution exists at $0.0055 < G <
0.0092$. Outside this region, the solution describes a uniform
system: the gas phase at the pumping less
 the lower boundary value and the condensed phase at the pumping greater upper boundary
 value. Upper part of the periodical distribution corresponds to condensed phase, the lower one corresponds to the gas phase.
 The size of the condensed phase increases with the change of the pumping from  $G_{c1}$ to $G_{c2}$.
  At $G>G_{c2}$ the state with uniform distribution of condensed phase emerges.
 \begin{figure}\centerline{\includegraphics[width=8.6cm]{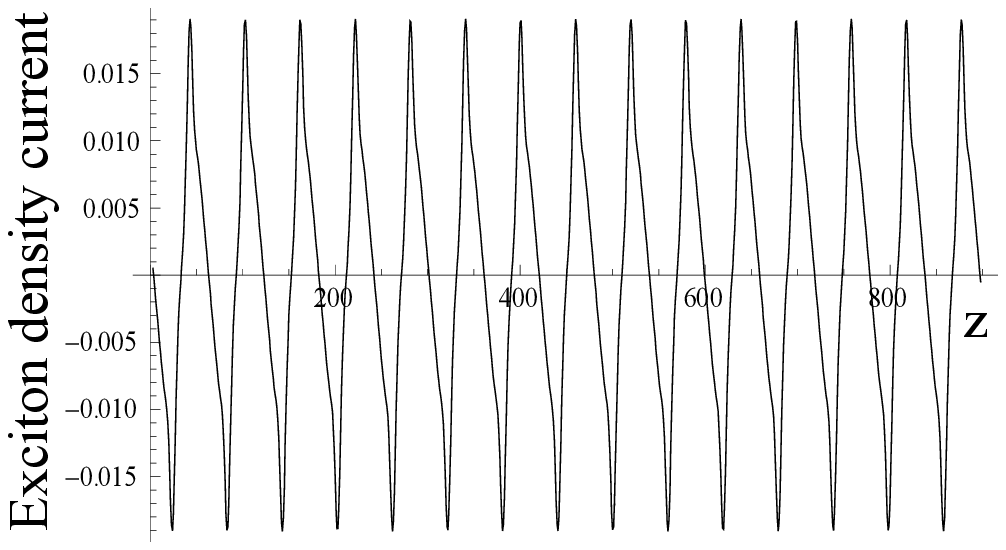}}\caption{
 The spatial dependence of the exciton current at  $G=0.008$,  $D_1=0.03$, $b=-1.9$.   }\end{figure}

  Fig.3 shows the spatial dependence of the exciton current
 calculated by the formula (\ref{eq16}). The current equals zero in
 the centers of the condensed and gas phases and it has a maximum in the
 region of a transition from the condensed phase to the gas
 phase.

 Let us do some estimations. It is seen from Fig.2, that the
 exciton density in the condensed phase redouble approximately in
 comparison with the density in the gas phase. The results for the
 currents in Fig.3 is presented in dimensionless units:
 $\tilde{j}=j/j_o$, where $j_o=n_ou_o$,
 $u_o=(\tau_{sc}n_oa)/(m\xi)$ is the unit of the velocity. The exciton density
 presented in Fig.2 in dimensionless units ($\tilde{n}=n/n_o$). It is seen from Fig.2 that$\tilde{n}\sim 1$,
 and the magnitude of $n$ has an order on $n_o$. So, for
  estimations we may suppose that $n_oa$ corresponds to  the shift
 of the luminescence line with increasing the exciton density,
 also  the magnitude of $\xi$ has the order of the size of the
 condensed phase. For the following magnitudes of parameters $\tau_{sc}=10^{-11}\textmd{s}, n_oa=2\cdot10^{-3}\textmd{eV},
  m=2\cdot10^{-28}\textmd{g},
   \xi=2\cdot 10^{-4} \textmd{cm}$, we obtain $u_o\sim 10^6 \textmd{cm/s}$. According
  to calculations (see Fig.3) the magnitudes of the current and
  the velocity are   in two orders of the values less than their
  unit. So, at chosen parameters, the maximal amplitude of the velocity of the exciton density flux
   in the dissipative structure in the double quantum well  equals
   $10^{4}\textmd{cm/s}$. In order to verify the fulfilment of the condition ($\ref{eq14b}$), let us suppose that
    ($\partial u_i)/(\partial x_k\sim u/l$), where $l$ is
    the period of the structure. It  follows from experiments \cite{5,9} that $l\sim (5\div
    10)\mu
    \textmd{m})$. Using these date we obtain that the condition
    ($\ref{eq14b}$) is satisfied very good. This condition is violated  at $\tau_{sc}\geq 10^{-9}\textmd{s}$. It is the very large value.
    The calculation using uncertainty principle from the
    band width of the narrow line ($2\cdot 10^{-4}\textmd{eV}$)\cite{2,3}
    gives more less the magnitude $3\cdot 10^{-12}\textmd{s}$.
    Therefore, the formation
    of nonuniform exciton dissipative structures  in the double
    quantum well occurs by the diffusion movement of excitons.
For a proof of the main equation (\ref{eq17}) the last term in Eq.
(\ref{eq9}), which describes the loss of the momentum due to
scattering on defects  and phonons, is of importance.  Just this
term describes the processes, which cause a decay of the exciton
flux. From the viewpoint  of the possibility of the appearance of
superfluidity, the situation for excitons is more complicated than
that for  the liquid helium and for the atoms of alkali metals at
ultralow temperatures.
 In the last systems, the phonons (moving of particles) are intrinsic
 compound part of the system spectrum, the interaction between
 phonons (particles) are the interaction between of atoms of the system and does not cause the change
 of the complete momentum of the system and  its moving as whole. Phonons and defects
 for excitons are  external subsystems, which brake the exciton moving. Therefore,
  for the creation of exciton superfluidity it is needed, that the value of $\tau _{sc}
$ growths significantly. It is possible for exciton polaritons,
which weekly interact with phonons; and there is a certain
experimental evidence on an observation of the polariton
condensation \cite{40}. For the indirect excitons the critical
temperature of the superfluid transition is strongly lowered by
inhomogeneities \cite{40a,40b}. So, the question about the
possibility of the superfluidity existence for the indirect
excitons on the base AlGaAs system is open.

  Thus, the peculiarities, that are observed at large densities of the indirect excitons,
  may be explained by the phase transitions in the system of the particles with attractive
   interactions and the finite value of the lifetime without an involvement of the Bose-Einstein condensation.

\section{Excitonic autosolitons}

As it was shown, at  $n < n_{c1} (G < G_{c1} )$ the uniform
solution of Eq. (\ref{eq19}) is stable. But, at some limits of a
pumping at $G < G_{c1}$ there exists a stationary solution for the
exciton density distribution localized in a space.
 For example, with the  parameters used for calculations of the exciton distribution in Fig.3,
  the threshold value of the pumping equals   $G_{c1} = $0.0055;
  but, at a steady-state pumping there is the spatial nonuniform solution of the equation
(\ref{eq19}) at $G<G_{c1}$ in the form of an isolated spike. It
may be obtained solving Eq.(\ref{eq19}) at the pumping, which
consist of a constant value  $G_0 $ and an additional pulse
 $dG$ with the maxima in the some point of the space and in  the time
 moment
\begin{equation}
\label{eq22}
dG = s\,\exp [ - w (z - z_0)^2]\,\exp [ - p (t - t_0
)^2]
\end{equation}
\noindent where $s,\,w ,\,p $ are parameters. The formula
(\ref{eq22}) describes a pulse of the pumping, which acts during
 some time interval  with the maximum in the point  $z_0$.

The solution of Eq.(\ref{eq19}) obtained  under an application of
the addition pulse (\ref{eq22})
 in the region $z_0=L/2$ has at $t \to \infty $ the form presented in Fig.4.
\begin{figure}\centerline{\includegraphics[width=8.6cm]{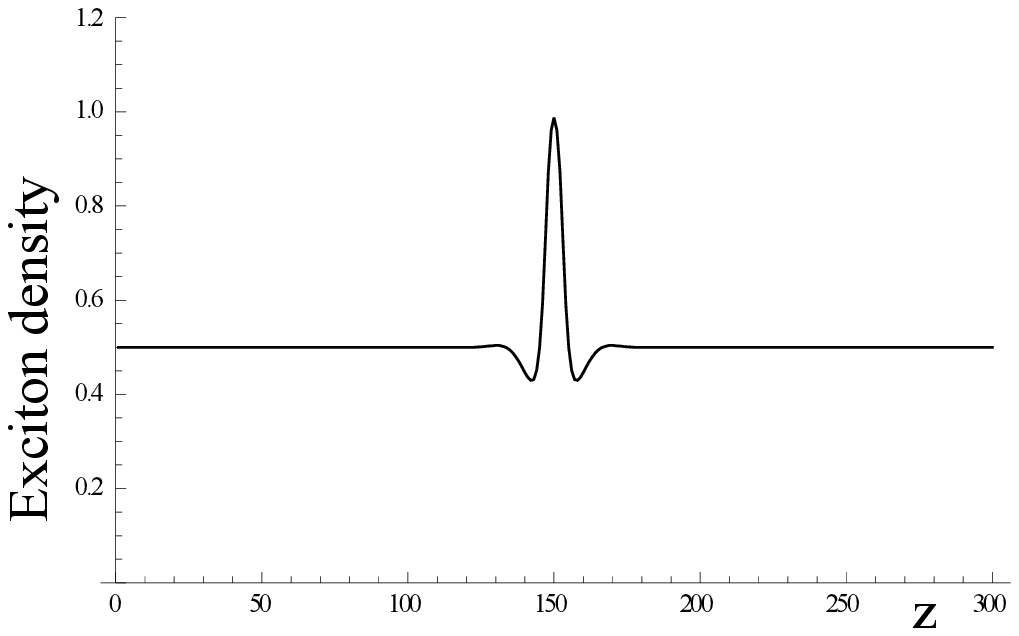}}\caption{
The spatial dependence of the exciton density of the excitonic
autosoliton at the pumping
 $G=0.005<G_{c1}$; $D_1=0.03$, $b=-1.9$. }\end{figure}
The solution exists at
 $t \to \infty $, i.e. at the times, when the action of the addition pulse is absent already.
  The shape of the spike  $n(z)$ does not depend on parameters $s,\;w,\;p$, except  cases,
  when at least one of these parameters tends to zero and becomes less some value.
  In addition, the solution in the form, presented in Fig.4, arises also, if the additional pulse is absent,
   but  there is some distribution of the exciton density in the initial time $t=0$:
\begin{equation}
\label{eq23}
n(z,0) = s_0\exp ( - w(z - z_0 )^2 ).
\end{equation}
Fig.5 shows the distribution of exciton current in the vicinity of
the localized solution. The current changes the sign in the center
of the localized state of the density.
\begin{figure}\centerline{\includegraphics[width=8.6cm]{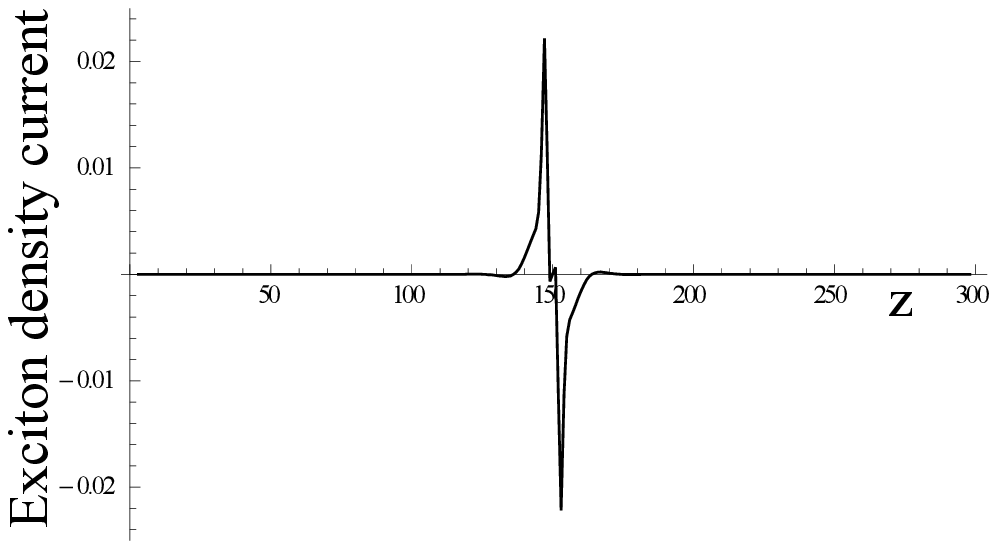}}\caption{
The spatial dependence of the exciton current in the localized
state at the pumping
 $G=0.005<G_{c1}$; $D_1=0.03$, $b=-1.9$. }\end{figure}

We have verified by direct calculations, that the solution
presented in Fig.4 in the form of a localized distribution of the
density is stable. We  call the state, that describes this
solution localized in space, by  excitonic autosoliton. The
spatial dependence of the exciton density will be designated by
$n_{as} (z)$. The autosolitons exist in the some regions of the
pumping $G_{cas} < G < G_{c1} $. At chosen parameters in Fig.4 the
autosolitons arise under conditions $0.003 < G < 0.0055$. The
solutions exist in the form of the autosoliton side by side with
the uniform solutions. The lower boundary $G_{cas}$ depends on
exciton parameters, particulary, on the exciton lifetime.

Excitonic autosolitons correspond to solitary solutions of
nonlinear equations for the excitons  (\ref{eq19}). The
name``\textbf{auto}'' is introduced according to \cite{42}; it
underlines, that the solitary solutions arise in the dissipative
system in a contrast to "solitons", which appear in conservative
systems. The obtained here autosolitons correspond to the "static
autosolitons" according to the classification \cite{42}. The
solutions in the form of the autosolitons are degenerated:  if
there is a solitary solution $n_{as} (z)$,then $n_{as} (z - z_0 )$
will be also solution at the arbitrary  $z_0 $ (in the infinite
medium). But, if there is an external field in a system, which
creates a spatial dependent additional potential for excitons, the
solitary excitation moves. Thus, at linear spatial dependence of
the additional potential energy in the formula of the free energy
(\ref{eq18}) the term $\delta V = - dz$ should be added. In this
case the equation
 (\ref{eq19}) has the solution in the form of autowaves  $n_{as} (z
- vt)$, where $v$  is the velocity of the autowave. In the region,
in which the periodic solution of the exciton density takes place
($G(c1) < G < G(c2)$), such autowaves were investigated in the
work \cite{Sug}.

Localized solutions exist also in the some region  at the pumping
greater the value, at which the periodical structure arises
($G>G(c2)$). The dependence of the exciton density may be obtained
from Eq.(\ref{eq19}) choosing an additional pumping pulse in the
form (\ref{eq22}), but at $a < 0$. An example of such solution is
presented in Fig. 6. These structures  appear in the form of a
dip, and can be called "dark autosolitons" by analogy with the
soliton's terminology.
\begin{figure}\centerline{\includegraphics[width=8.6cm]{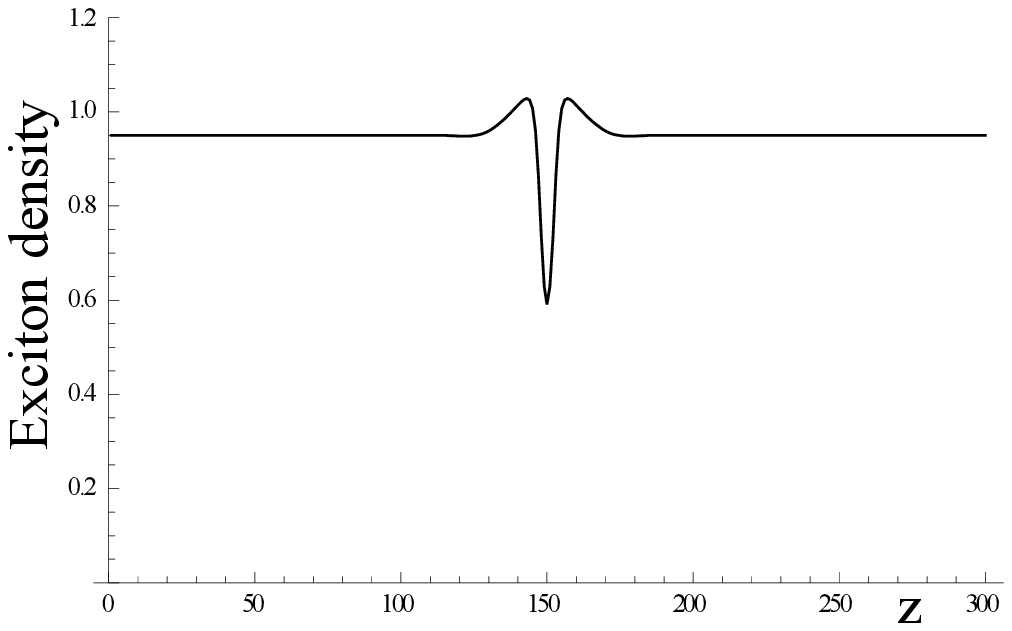}}\caption{
The spatial dependence of the exciton density of the dark
autosoliton at $G=0.0095>G_{c2}$; $D_1=0.03$, $b=-1.9$  .
}\end{figure}

To explain  the appearance of the autosoliton-type
solution we recall that the phase transition are investigated in
the paper. As is known, there exists the region
 between spinodal and  binodal, in which  the creation of a nucleus of new phase is needed for phase transition.
  The size of the nucleus  should exceed some critical value.  The obtained critical value of $n_c$ (Eqs.(\ref{eq14ab},\ref{eq14ac})) is
   based under the consideration of the small fluctuations and corresponds really
 to the boundary of the spinodal region corrected by non-equilibrium state of the system.
The autosoliton arises at pulses larger some critical
    value.
  So, the appearance  of autosolitons
    corresponds to the appearance of the nucleus outside the spinodal boundary for phase transitions of the stable
    particles.
      A size of
the new phase of the stable particles  increases with time, while
the distribution density of the unstable particles (excitons) does
not depend on the time (at a steady-state pumping). It explains
the existence of the localized states of exciton distribution
outside the region of the instability of an uniform distribution.
It should be noted, that in the phase transition approach of
nucleation-growth with an generation on unstable particles
\cite{19,20,21} the existence  of
 the condensed phase islands outside the spinodal region is taking
 into account automatically, because  this approach consider fluctuations, which do not presented in the equation  (\ref{eq19}).

 The model of the nucleation-growth,
  which we used under the investigation of the exciton condensation in \cite{19,20,21}, is close
  to the methods, used in \cite{43,44,45} under the investigation of electron-hole droplets
  in germanium and silicon. But in contrast to the works  \cite{43,44,45} we took into
  account the correlation in the spatial positions of the droplets (the islands in the two-dimensional case),
   that allows to describe the different structures, mutual positions of  of islands, their
   concentration, the formation of the periodicity
    and other properties, which are observed in quantum wells on the base of  the AlGaAs system.

\section{Conclusions}
In the work, the several problems of the theory of indirect
exciton condensation in quantum well on the base AlGaAs crystal
are considered.

 The hydrodynamic equations for excitons were build with taking into account a possibility of the condensed phase formation,
 the finite lifetime of excitons, the scattering of excitons by
 defects. The analysis of the equations was fulfilled for the diffusive
 exciton flow. The equations explain many  spatial excitonic structures obtained under experimental   investigations
  of indirect excitons in quantum wells on the base AlGaAs without
  an involvement of the Bose-Einstein condensation of excitons.

  Also, solutions of the equations in a form of  solitary localized states (the bright and dark exciton autosolitons)
 were found. They exist side by side with the steady-state exciton density state and may be excited by the additional
 pulse greater some threshold value. In the model of the nucleation-growth these solutions correspond to the
 nuclei
 of the condensed phase in a gas phase (the bright autosolitons)
  and to the nuclei of the gas phase in the condensed phase (the dark autosolitons).

 It was shown, that the results of the experiment
\cite{30}, in which has been obtained, that the emission spectrum
 from the condensed phase region is shifted to shortwave side in a comparison of the emission spectra from the region of the gas
 phase, do not contradict to the model with an attractive interaction between excitons.  For the explanation it was taking into
  account, that a formation of the emission spectrum occurs by both  free  and   trapped excitons.

\appendix\section{Distribution of excitons between localized and delocalized
states} According to \cite{30}  the frequency of the emission from
the islands on the ring around the laser spot is higher, than the
frequency of emission from  the region between islands. The
authors made the conclusion \cite{30}, that interaction between
excitons is repulsive, and, therefore, the formation of the
condensed phase by attractive interaction is impossible.  It
contradicts the main assumption of the works
\cite{19,20,21,22,23}, though these works explain many
experiments. Now we remove this contradiction, taking into account
the presence of localized excitons.

 Residual donors and acceptors, defects,
inhomogeneous thickness of the wells create an accidental
fluctuating potential, which may be the reason of appearance of
the localized levels. Till now the explanation of the creation of
the localized states is not determined definitely, but their
existence is confirmed by the presence of an emission in the
region of the frequencies less the frequency of the exciton band
emission and broadening of exciton lines. At the low temperature
and at the small pumping the main part of the band consists of the
emission from defect centers, the part of the exciton emission
growths with increasing pumping.  Let us consider the relation
between the contribution to the emission band intensity from free
excitons and the excitons localized on defects. Since the defect
structure of the samples depends on their preparation, a solution
of this problem can not be  solved in general. We shall use some
approximations.

We shall consider the energy distribution  of electron-hole pairs
at steady-state irradiation. Such pair may be the delocalized
exciton in the exciton band and an electron and hole localized
near a defect or in the region with a modified thickness of the
well. For small density of the excitation (the electron-hole
pairs) the interaction between them may be neglected. Due to
long-range character the dipole-dipole interaction, which appear
with increasing the excitation density, gives  an identical shift
of delocalized and localized levels. It means that this
interaction do not influence on mutual distribution of the
localized and delocalized states. We shall suggest that
 the localized states are saturable,
namely, every center may capture a restricted number of
excitations. In our calculations we shall assume  that only single
excitation may be localized on the defect. Another excitations are
or absent or have very low binding energy and are unstable. The
dependence of a density of localized states on energy was chosen
in  the exponential form, namely $\rho (E) = \alpha N_l exp(
\alpha E)$, where $N_l $ is the density of the defect centers, $E$
is the depth of the trap level. The exciton states (free and
localized) are distributed on levels after a creation of electrons
and holes by an external irradiation and their subsequent
recombination and relaxation. Because the time of the relaxation
is much less than the exciton lifetime,  the distribution of
excitation between free and localized states corresponds to to a
state of thermodynamical equilibrium. In considered model we
should obtain the distribution of electron-hole pairs,  a
population of which on a single level may be changed from zero to
infinity for $E>0$ (the free exciton states) and from zero to one
for $E<0$ (the localized states). Formally, in considered system
the free excitons have Bose-Einstein statistic and localized
excitations obey Fermi-Dirac statistic. At small exciton density
Bose-Einstein
 and Boltsmann statistics give the similar results for the free excitons,
  but the application of Fermi-Dirac statistic for localized states with single level for one trap is important.
   The equation for the energy distribution
may be find from minima of large canonical distribution
\begin{equation}
\label{eqa} w(n_k,n_i)=\exp\left(\frac{\Omega+N\mu-E}{\kappa
T}\right),
\end{equation}

\noindent
 where $N=\Sigma_in_i+\Sigma_kn_k$,
$E=\Sigma_in_iE_i+\Sigma_{k,l}n_kE_k$, $n_i=0,1$,
$n_k=0,1,...\infty$, $k$ is the wave vector of the exciton, $l$
designates the singular levels. $\mu$ is the exciton chemical
potential.

The distribution of excitons over free and localized level is
determined from minimum of the functional (\ref{eqa}). As the
result we obtain the following conditions for the mean values of
the  free exciton density $n$ and the density of the localized
states $n_L$
\begin{equation} \label{eqb} n_{ex}=\frac{g \nu}{4\pi
E_{ex}a_{ex}^2}\int_0^\infty\frac{dE}{\exp(\frac{E-\mu}{\kappa
T})-1} ,
\end{equation}
\begin{equation} \label{eqc} n_L=\alpha N_l
\int_{-\infty}^0\frac{\exp(\alpha E) dE}{\exp(\frac{E-\mu}{\kappa
T})+1} ,
\end{equation}
\noindent where $a_{ex}=(\hbar^2\varepsilon)/(\mu_{ex} e^2)$ and
$E_{ex}=(\mu_{ex} e^4)/(2\varepsilon^2 \hbar^2)$ are the radius
and the energy   of the exciton in the ground state in bulk
material,  $g=4$, $\mu_{ex}$ is the reduced mass of the exciton,
$\nu$ is the ratio of the reduced and the total mass of the
exciton. The chemical potential $\mu$ is determined from condition
\begin{equation} \label{eq6} n_L + n = G\tau_{ex},
\end{equation}
\noindent where  $G\tau_{ex}$
 is the whole number of excitation (free and localized) per unit surface.

The dependence of the distribution of the free and the localized
exciton on pumping is presented in Fig.7 as function of whole
number of the excitation presented in  units of $1/a_{ex}^2$
\begin{figure}\centerline{\includegraphics[width=8.6cm]{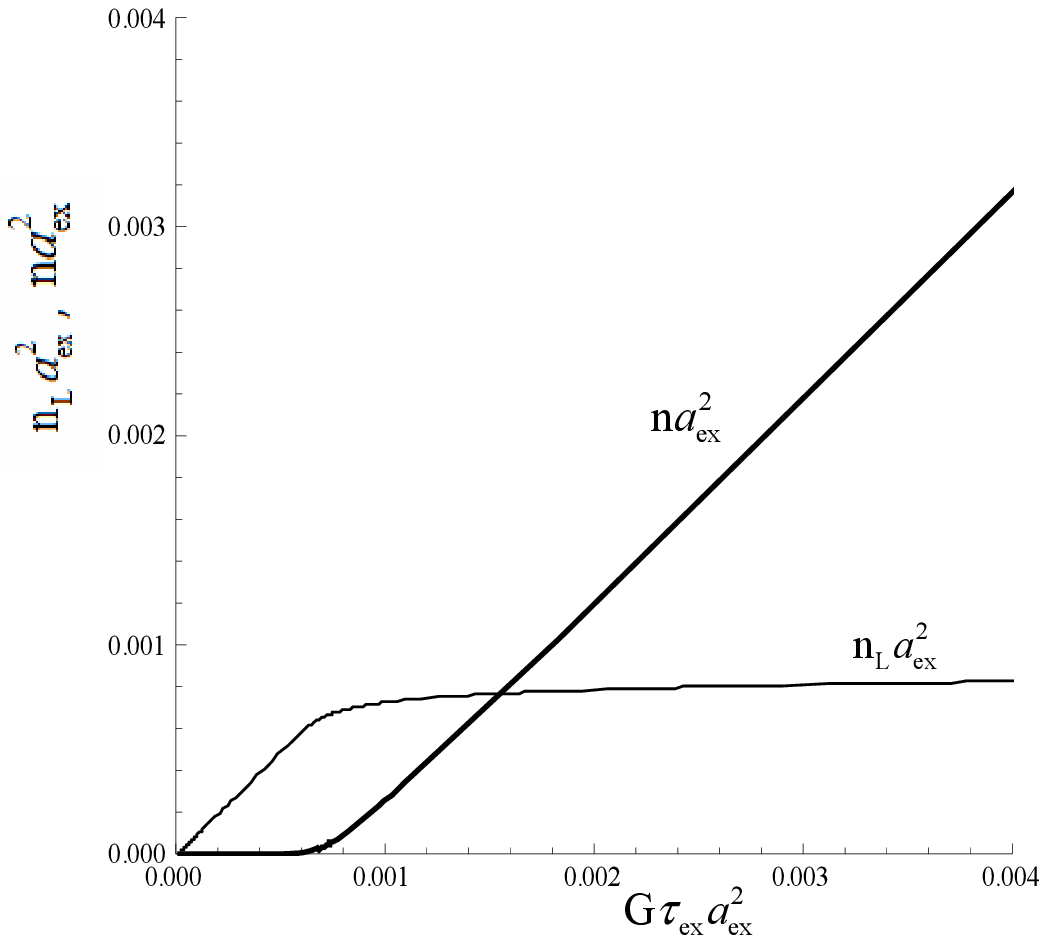}}\caption{
The dependence of the density of the free (thick line) and trapped
(thin line) excitons on the pumping. The parameters of the system:
$T=2K,  N_l=0.001/a_{ex}^2,
 \alpha=300 (\textmd{eV})^{-1}$    }\end{figure}
Let us the exciton radius equals $10\textmd{nm}$. Then the
concentration of the traps and the width of the distribution of
trap levels, chosen under calculations of Fig.7, have the order of
$10^9 \textmd{cm}^{-2}$  and $0.003 \textmd{eV}$, correspondingly.

As it is seen from Fig.7 that the number of the localized
excitations exceeds at small pumping the number of free excitons
and the emission band should be determined by the emission from
the traps. With increasing   pumping the occupation of the trap
levels become saturated. For chosen parameters  the concentration
of excitations at the saturation is a value of the order of $10^9
\textmd{cm}^{-2}$. Simultaneously with the saturation of the
localized levels the exciton density growths. As the result, the
shortwave part of the emission band should be increased with
increasing pumping. When the exciton density
 becomes great, the collective exciton effects  begin to manifest themselves. The equations (\ref{eqb},\ref{eqc}) do not
 take into account the interactions between the excitations, and special models and theories are needed
  for  descriptions of collective effects.
 The appearance of a narrow line was observed in \cite{2} with increasing pumping on the shortwave
part on  the exciton emission band. Simultaneously, the patterns
arise in the emission spectra. The narrow line appeared after the
localized states becomes occupied. According to \cite{2} this line
is connected with the exciton Bose-Einstain condensation.
According to our model \cite{19}, the islands of condensed phase
arise, if the exciton density become higher some threshold value,
and the narrow band corresponds to the condensed phase, caused by
the attractive interaction between excitons.
\begin{figure}\centerline{\includegraphics[width=8.6cm]{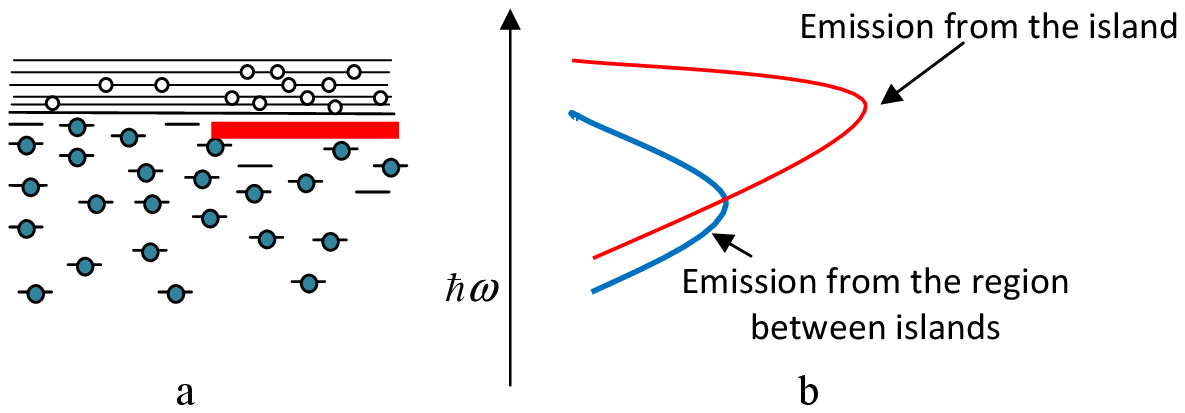}}\caption{
The distribution of excitations in the traps and in the states of
exciton band. The thick line in fig.8a corresponds to the energy
per single exciton in the condensed phase. On the right (fig.8b)
the upper line describes the whole  emission from the island (the
emission of both the condemned phase and trapped excitons), the
low line describes the emission of the trapped excitons.
}\end{figure} The energy per single exciton in the condensed phase
is less than the energy of free excitons (the thick line in
Fig.8), but the gain of energy under condensation is less than the
whole bandwidth, which are formed by the localized and delocalized
states. The gain of energy is significantly less than the binding
energy of the  exciton  to an electron -hole drop in silicon and
germanium. According to \cite{3,4} the narrow band is shifted to
the red side with increasing pumping in the value less than 0.5
meV, while the whole bandwidth has the order of 2 \textmd{meV}.
So, the energy of photons emitted from the islands of condensed
phase is higher than the energy of photons emitted by traps. The
excitons can not leave the condensed phase (the islands) and move
to the traps (to the states with lower energy) since the levels of
the traps are occupied already. Thus, the emission frequency of
the condensed phase is larger, than the emission frequency of the
gas phase, \textit{even at the attractive interaction between
excitons}. It may be reason of obtained in \cite{30} results in
which the maximum of the frequency of the emission from the
islands is higher than the maximum frequency from the regions
between the islands.

The qualitative results coincide with the results obtained from
the solution of kinetics equations in \cite{SugUJP} for level
distributions. The similar behavior of the distribution of free
and trapped excitons takes place for an other  energy dependence
of the density of localized states. We confirmed the results for
the gaussian density distribution.

 The results may be applied for explanation of
intensity and temperature dependence of the exciton emission of
dipolar excitons in InGaAs coupled double quantum wells
\cite{Schin}. The authors observed with increasing pumping the
growth of the shortwave side of the emission band, the narrowing
of the band. They obtained that the shortwave side of the band is
very sharp. These results  may be explained by suggestion, that
the lower part of the band is formed by localized states. After
the saturation of the localized states with increasing pumping,
the excitations begin to occupy the levels of  free excitons.
Since the density of exciton state is much greater than the
density of defect states the shortwave edge of the band is sharp.
The exciton condensed phase may not arise in the system
investigated in \cite{Schin} as the distance between wells
($17\textmd{nm}$) is greater  than the distance in the system
investigated by the Timofeev's group \cite{2}($13\textmd{nm}$), so
the repulsive dipole-dipole interaction between excitons in
\cite{Schin}  may be larger than the attractive interaction.

 It should be noted, that in the
work \cite{5} the band of the emission from the condensed phase is
wider than the narrow band that was observed in the work \cite{4},
where the another method of the creation of excitons was applied.
Maybe, it is related to the fact, that in the work \cite{5},
excitons were created in the region of the p-n transition:  the
electrons approach  this region from one side, and the holes move
to the region from other side. The excitons are situated in a
space between the regions rich by electrons from one side, and by
holes from other side. To combine into excitons, electrons and
holes should surely go through the region of the condensed phase.
Therefore, this region contains the charges, which create an
electric field and cause the broadening of the emission band.

\end{document}